\documentclass[aps, prl, 10pt, superscriptaddress, twocolumn]{revtex4-1}
	\usepackage[bookmarks=false,linkcolor=blue,urlcolor=blue,colorlinks,citecolor=blue]{hyperref}

	\usepackage{graphicx}
	\usepackage{color}
	\usepackage{amsmath}
	\usepackage[normalem]{ulem}
	\usepackage[toc,page]{appendix}
	\usepackage{amsfonts}
	\usepackage{amssymb}

\newcommand{\be}{\begin{equation}}
\newcommand{\ee}{\end{equation}}
\newcommand{\p}{\partial}

\newcommand{\la}{\label}
\newcommand{\bea}{\begin{eqnarray}}
\newcommand{\eea}{\end{eqnarray}}

\definecolor{cardinal}{rgb}{0.6,0,0}
\definecolor{darkgreen}{rgb}{0,0.4,0}
\definecolor{golden}{rgb}{0.92, 0.7, 0}
\definecolor{midnight}{rgb}{0, 0, 0.5}
\definecolor{darkblue}{rgb}{0, 0, 0.7}

\begin{document}

\title{
	A Duality Between $U(1)$ Haah Code and $3$D Smectic A Phase
	 \vskip 20pt
	 }

\author{Andrey Gromov}
\affiliation{Brown Theoretical Physics Center \& Department of Physics, Brown University, Providence, Rhode Island 02912, USA}

\date{\today}

%%%%%%%%%%%%%%%%%%%%%%%%%%%%%%%%%%%%%%%%%%%%%%%%%%%%%%%%%%%%%%%%%%%%%%%%%%%%%%%%%%%%%%%%%%%%%%%
\begin{abstract}
We describe a duality between multipole gauge theories and spatially ordered phases. Our main example is a duality between the multipole gauge theory description of the $U(1)$ Haah code and smectic A phase in three spatial dimensions. We show how multipole symmetries restrict the mobility of dislocations and disclinations in the smectic A phase. We also exhibit a $2$D version of the  duality.
\end{abstract}

%%%%%%%%%%%%%%%%%%%%%%%%%%%%%%%%%%%%%%%%%%%%%%%%%%%%%%%%%%%%%%%%%%%%%%%%%%%%%%%%%%%%%%%%%%%%%%%

\maketitle

%%%%%%%%%%%%%%%%%%%%%%%%%%%%%%%%%%%%%%%%%%%%%%%%%%%%%%%%%%%%%%%%%%%%%%%%%%%%%%%%%%%%%%%%%%%%%%%

%%%%%%%%%%%%%%%%%%%%
\paragraph{Introduction.---} 
%%%%%%%%%%%%%%%%%%%%
Fracton phases of matter are characterized by the presence of immobile local excitations. One systematic approach to these phases is based on tensor \cite{xu2006novel,xu2010emergent,rasmussen2016stable, pretko2017subdimensional, pretko2017generalized} and multipole gauge theories (MGT) \cite{gromov2018towards, bulmash2018generalized}. To describe gapped fracton phases these theories have to be discretized on a lattice and gapped via Higgs mechanism \cite{BB2018Higgs, ma2018fracton}. It was known for a long time that ordinary elasticity is dual to symmetric tensor gauge theories (STGT) \cite{kleinert1982duality, kleinert1983dual, kleinert1983double}.  Lattice defects in quantum theory of elasticity share some common properties with fractons \cite{pretko2018fracton, gromov2019chiral, radzihovsky2020fractons, gromov2019duality, pretko2018symmetry, kumar2018symmetry, pretko2019crystal}. Namely, the restricted mobility: dislocations satisfy the glide constraint that forces them to move along their Burgers vector, while disclinations are immobile.

Certain fracton phases cannot be described by STGTs and require a more abstract type field theories, which were termed \emph{multipole} gauge theories \cite{gromov2018towards, bulmash2018generalized}. In particular, the ubiquitous Haah code \cite{haah2011local} as well as the Chamon code \cite{Chamon2005} are both related to such theories (as shown in \cite{gromov2018towards, bulmash2018generalized} and \cite{you2019fractonic} correspondingly). MGTs are very unusual Abelian gauge theories and their physical interpretation is not always clear. One puzzling property of such theories is that they often require a length scale in order to define an effective action, which indicates sensitivity to the UV physics. Generally, these theories are also anisotropic and can admit an exotic version of Lifshitz symmetry, with several dynamical critical exponents \cite{gromov2018towards}. 

In this Letter we show that a particular multipole gauge theory, namely the one related to the $U(1)$ Haah code is dual to a (quantum) $3$D smectic A phase \cite{de1993physics}, which was anticipated in \cite{gromov2018towards}. This duality shows that some multipole gauge theories can be understood in terms of familiar physical systems, that are naturally anisotropic and depend on an emergent length scale (namely, the separation between the layers in the smectic phase). We start with a brief introduction to the multipole gauge theories. Then we consider a warm up example -- a duality between $2$D smectic and a $2$D multipole gauge theory. This is followed by the formulation of the $U(1)$ Haah code. Next we explain how to map it onto the smectic A phase. We investigate symmetries, conservation laws and mobility of the defects in the dual theory. Conclusions based on symmetry analysis agree with what is known about smectic phases. 

%%%%%%%%%%%%%%%%%%%%
\paragraph{Multipole gauge theory.---} 
%%%%%%%%%%%%%%%%%%%%
We start with a review of the multipole gauge theories \cite{gromov2018towards} necessary to describe the $U(1)$ Haah code as well as the Chamon code. 
To define a multipole gauge theory, one needs a set of derivative operators
\be
D_{I} = q^i_I\p_i + q^{ij}_I\p_i \p_j + \ldots\,.
\ee
The coefficients $q^i_I, q^{ij}_I,\ldots$ are vectors and symmetric tensors and are \emph{dimensionful}. These differential operators essentially define the multipole gauge theory structure.

Given the derivative operators one introduces a set of electric fields and vector potentials
\be
E_I = \p_0 A_I - D_I \Phi\,,
\ee 
where $A_I$ is the ``vector'' potential and $\Phi$ is the scalar potential. The index $I=1,2,\ldots$ is \emph{not} a spatial or space-time index. Generally speaking, the fields $A_I$ and $E_I$ transform in an unpredictable way under rotations. Although in all studied examples they fall into representations of a discrete point group symmetry. The variables $A_I$ and $E_I$ are canonically conjugate to each other
\be
[A_I(x), E_J(x^\prime)] = i \delta(x-x^\prime)\,.
\ee 
The electric fields $E_I$ are invariant under the following gauge transformations
\be
\delta A_I = D_I \alpha\,, \qquad \delta \Phi = \dot{\alpha}\,.
\ee
These transformation laws are generated by the following Gauss law
\be
\sum_I D^\dag_I E_I = \rho\,,
\ee
where $D^\dag_I$ is defined via integrating $D_I$ by parts and $\rho$ is the density of charges that source the electric $E_I$ fields. The action is given by a generalized Maxwell form
\be
S =\int  \epsilon\sum_I E_I - \sum_{I\neq J} B_{IJ}\,,
\ee
where $B_{IJ} = D_IA_J - D_J A_I$ is the mangetic field \footnote{Definition of the magnetic field involves various subtleties in the cases when the invariant derivatives are not algebraically independent}.
This structure describes a vast landscape of Abelian, non-relativistic gauge theories, which remains largely unexplored. 

One way to classify the possible choices of $D_I$ is to note that the Gauss law implies a set of non-trivial conservations laws. Indeed, assuming the charge conservation law
\be\la{eq:conservation}
\partial_0 \rho + \sum_I D^\dag_I J_I =0
\ee
we find 
that 
\be
\la{eq:multipoles}
\partial_0 \int d^3x P_A(x) \rho =0\,,
\ee
for all polynomials that are annihilated by all derivative operators simultaneously
\be
\la{eq:poly}
D_I P_A(x) = 0\,.
\ee
Most fracton models fit into this construction \footnote{A notable misfit is the Haah's B code, which requires two conserved charges and, consequently, should correspond to a $U(1)\times U(1)$ gauge theory, which remains to be developed.}.

%%%%%%%%%%%%%%%%%%%%
\paragraph{Multipole gauge theory for a $2$D smectic.---}
%%%%%%%%%%%%%%%%%%%%
Before moving on to the $U(1)$ Haah code we pause to discuss a simple warm up case. We will show that a $2$D quantum smectic is dual to a multipole gauge theory. We start with the following action \cite{de1993physics} that describes a smectic phase at long distances
\be
S = \int \dot{\theta}^2 -   \epsilon^{-1}(\p_w\theta)^2 - \epsilon^{-1}(\lambda\p_u^2 \theta)^2\,,
\ee
where the layers are perpendicular to the $w$ direction and extend in $u$ direction. We remind the reader that smectic phases break spontaneously one out of two translation symmetries. They can be viewed as stacks of $1$D liquid phases separated by some distance, $d\approx \lambda$, in $w$ direction. We denote 
\be
D_1^\dag = \p_w\,, \qquad D_2^\dag = \lambda\p_u^2\,. 
\ee
In terms of these derivatives the action is
\be
S = \frac{1}{2}\int \dot{\theta}^2 -   \epsilon^{-1}(D_1^\dag\theta)^2 - \epsilon^{-1}(D_2^\dag \theta)^2\,.
\ee
We introduce auxiliary variables 
\be
S = \int P \dot\theta - \frac{P^2}{2} - (D_1^\dag\theta) T_1   - (D_2^\dag\theta) T_2 +\epsilon\frac{T_1^2}{2}+ \epsilon\frac{T_2^2}{ 2} 
\ee
Integrating out the phonon $\theta$ we find a constraint
\be
\dot{P} - D_1 T_1 - D_2 T_2 = 0\,.
\ee 
This equation is solved by 
\bea
&&T_I = \epsilon_{IJ}(\dot{A}_J - D_J \Phi) = \epsilon_{IJ}E_J\,, 
\\
&&P = B = \epsilon_{IJ}D_IA_J\,,
\eea
where $\epsilon_{IJ}$ is the Levi-Civita symbol. The gauge redundancy of the solution is 
\be\la{eq:2DMGT}
\delta A_I = D_I\alpha\,, \quad \delta \Phi = \dot\alpha
\ee
which is exactly a multipole gauge theory structure. The Gauss law that generates \eqref{eq:2DMGT} is given by
\be\la{eq:2DGauss}
D_I^\dag E_I =\rho\,.
\ee
The defect density $\rho$ is the density of smectic disclinations. The defect matter conserves the dipole moment in $u$ direction, which can be seen directly from \eqref{eq:2DGauss}. Disclination dipole extended in the $u$ direction is a dislocation with the Burgers vector in $w$ direction. The dislocations are completely mobile \cite{kleman2007soft}, whereas the disclinations are $1$D particles (also known as lineons) that can only move in the $w$ direction. The low energy phonon is described by the multipole gauge theory with the generalized Maxwell action
\be\la{eq:2DMGT}
S = \frac{1}{2}\int  \epsilon(E_1^2 +   E_2^2) - B^2\,.
\ee
There is a single mode with linear dispersion in $w$ direction and quadratic dispersion in $u$ direction.

%%%%%%%%%%%%%%%%%%%%
\paragraph{$U(1)$ Haah code.---} 
%%%%%%%%%%%%%%%%%%%%
The Haah code was originally defined via a commuting projector Hamiltonian \cite{haah2011local, haah2013commuting}. Later it was realized that the physics of the groundstates and topologically non-trivial excitations can be reproduced via a $\mathbb{Z}_2$ multipole gauge theory which also has a natural $U(1)$ analogue  \cite{gromov2018towards, bulmash2018generalized}. We refer to the corresponding $U(1)$ MGT as $U(1)$ Haah code. 

 The gauge theory is defined by the following set of derivatives
\be
D_1 = \p_x + \p_y + \p_z\,, \qquad D_2 = a(\p_x \p_y + \p_x\p_z + \p_y \p_z)\,,
\ee
where $a$ is a constant of dimension of length. A more general version of the gauge theory was defined in \cite{gromov2018towards} and includes one additional differential operator. To simplify the presentation we have opted to exclude it. This way there are no relations between the  derivatives $D_I$ and there is a unique magnetic field. In this gauge theory the charges can only appear in two configurations, which establishes the connection to the Haah code upon condensation of charge $2$ particles  \cite{gromov2018towards, bulmash2018generalized}. 

There are two gauge fields $A_1, A_2$ and two electric fields satisfying the Gauss law
\be\la{eq:Gausslaw}
D^\dag_1E_1 + D^\dag_2E_2 = \rho\,.
\ee
It is convenient to introduce another set of variables
\bea
&&w=\frac{1}{\sqrt{3}}(x+y+z)\,,
\\
 &&u=\frac{1}{\sqrt{2}}(x+y)\,,\qquad v=\frac{1}{\sqrt{6}}(x+y-2z)\,.
\eea
In these variables the derivatives take a particularly nice form
\be
D_1= \sqrt{3}\p_w\,,\qquad D_2 = a(\p^2_w - \frac{1}{2} \Delta_{u,v})\,,
\ee
where $\Delta_{u,v}$ is a $2$D Laplace operator in the $u-v$ plane.

The polynomials satisfying \eqref{eq:poly} were found in \cite{gromov2018towards}. For the present purpose it is more palatable to express these polynomials in terms of $(u,v,w)$ variables
\footnote{In terms of the original variables we have $R_1 = x-y\,, R_2 = x+y-2z\,, R_3 = R_1\cdot R_2\,, R_4 = -\frac{1}{4}(R_1-R_2)\cdot(3R_1+R_2)$}
\bea
P_1 = u\,,\quad P_2 = w\,,\quad P_3 = uv\,,\quad P_4 = u^2 - v^2\,.
\eea
Note that $(P_1,P_2)$ form a vector representation, while $(P_3,P_4)$ form a symmetric traceless representation of $SO(2)$.
Corresponding conserved quantities are given by \eqref{eq:multipoles}. These conservation laws imply that charges can only appear in certain geometric patterns with vanishing multipole moments \eqref{eq:multipoles}. The Gauss law is consistent with these conservation laws by construction. One can also observe by inspection that there is an infinite set of conserved charges 
\be\la{eq:sliding}
Q[f] = \int du dv \rho(u,v,w) f(u,v)\,,
\ee
where $f(u,v)$ is a harmonic function. These conservation laws will be explained shortly.

Finally, the effective action is given by
\be\la{eq:HaahGauge}
S = \int \epsilon(E_1^2 + E_2^2) -  B^2\,,
\ee
where $B=D_1A_2 - D_2 A_1$. Next we perform a duality transformation and interpret the dual theory as a theory of a quantum $3$D smectic.

%%%%%%%%%%%%%%%%%%%%
\paragraph{Duality transformation.---} 
%%%%%%%%%%%%%%%%%%%%
The dual variables are introduced via solving the Gauss law \eqref{eq:Gausslaw}. A generic electric field satisfying the Gauss law can be written as
\be\la{eq:Efield}
E_I = \epsilon_{IJ} D^\dag_J \theta\,, \qquad B = \dot{\theta}
\ee
where we have used $\delta_{IJ}$ to contract the abstract indices $I,J,\ldots$.
In terms of $\theta$ the action takes form
\be
S =\int  \dot{\theta}^2 - \epsilon (D_1\theta)^2 - \epsilon(D_2 \theta)^2\,.
\ee
The action is more transparent in the $(u,v,w)$ variables
\be
S = \int \dot{\theta}^2 - 3\epsilon(\p_w \theta)^2 - \epsilon a^2\left((\p_w^2 - \frac{1}{2} \Delta_{u,v})\theta\right)^2\,,
\ee
which is the Lagrangian for a $3$D (quantum) smectic in harmonic approximation. To make the comparison more explicit we re-arrange terms
\bea \nonumber
S =\int  \Big\{\dot{\theta}^2 &-& 3\epsilon (\p_w \theta)^2 - \frac{\epsilon a^2}{4}\left( \Delta_{u,v}\theta\right)^2 \Big\}
\\ \la{eq:smectic}
&-& \epsilon a^2 (\p_w^2 \theta)^2 + \epsilon a^2 (\p_w^2 \theta) (\Delta_{u,v}\theta)\,,
\eea
which is precisely the effective Lagrangian for a smectic A phase. The corresponding Hamiltonian can found in \cite{de1993physics}. The last two terms are less important for the long wavelength physics of smectics \cite{de1993physics}. Moreover, there are no symmetry constraints that enforce any particular value for the coefficient in front of $\p_w^2\theta$. Thus we can take it to zero without losing any essential properties.

We emphasize that from the smectic A point of view it is \emph{essential} that a length scale appears in the action explicitly. This length scale, $\epsilon^\frac{1}{2} a$, is of the order of the distance between the layers \cite{de1993physics}. It does not make sense to consider a limit $a\rightarrow 0$. 

Note that the actions \eqref{eq:2DMGT} and \eqref{eq:HaahGauge} are identical, with the same number of degrees of freedom and gauge fields. The real difference is in the gauge structure enforced on these fields, \emph{i.e.} in the explicit form of the derivatives $D_I$. This happens because in the smectic A phase the number of degrees of freedom is independent of the number of spatial dimensions: it is always a single phonon.

%%%%%%%%%%%%%%%%%%%%
\paragraph{Some properties of the dual theory.---} 
%%%%%%%%%%%%%%%%%%%%

The dual theory has a global symmetry under shifting $\theta$ by a constant. The conservation law is given precisely by \eqref{eq:conservation}. In the context of smectics, $\theta$ is a phonon along the (only) crystalline direction.  Consequently the shift $\theta \rightarrow \theta + c$ is a translation along the $w$ direction and the corresponding conservation law is conservation of momentum along $w$ direction, $P_w = \int p_w$. 

Next, we have  ``dipole'' symmetries  
\be\la{eq:smecticdipole}
\delta \theta = c_1 u + c_2 v\,.
\ee
These symmetries lead to a conservation law of the dipole moment of $p_w$
\be
D_u = \int u p_w\,, \quad D_v = \int v p_w\,.
\ee 
These dipole moments are just the components of the angular momentum $M_{uw}$ and $M_{vw}$ ($p_u$ and $p_v$ can be taken to be $0$). Indeed, this interpretation is particularly clear if we note that \eqref{eq:smecticdipole} is a rotation in the target space.

Discussion of the quadratic symmetries is more subtle. Indeed, the harmonic action is made exclusively from the derivatives $D_1$ and $D_2$. However, both of these derivatives annihilate \emph{any} harmonic function of $u$ and $v$, leading to an infinite set of conserved quantities discussed above \eqref{eq:sliding}. Geometrically these conservation laws state that bending all layers simultaneously costs no energy. These conservation laws are usually absent in more accurate theories of smectics because they are broken by anharmonic terms \cite{grinstein1981anharmonic} as well as by the boundary conditions. Given that conservation of the components of the quadrupole tensor is essential in the physics of $U(1)$ Haah code (because it specifies the form of $D_2$), it is likely that the correspondence with smectic A only holds at the quadratic level, while the non-Gaussian corrections will be \emph{different}.

%%%%%%%%%%%%%%%%%%%%
\paragraph{Defects.---} 
%%%%%%%%%%%%%%%%%%%%

Under the duality transformation the defects in smectic A phase will map to the charges coupled to the $U(1)$ Haah code. The defects in the smectic A phase are dislocations with the Burgers vector pointing in the $w$ direction. The multipole gauge theory naturally couples to the following matter Lagrangian 
\be\la{eq:discl3D}
\mathcal L = \dot{\Phi}^\star \dot{\Phi} + \alpha |(D_1 -i A_1)\Phi|^2 + \beta |D[\Phi,\Phi]|^2\,,
\ee
where
\bea
D[\Phi,\Phi] &=& q^{ij}\left( \partial_i\Phi\partial_j \Phi - \Phi \partial_i \partial_j \Phi\right) - iA_2\,,
\\
 q^{ij} &=& \frac{1}{2} \begin{pmatrix} 
			0&1&1
			\\
			1&0&1
			\\
			1&1&0
			\end{pmatrix}\,.
\eea
The form of the matrix $q^{ij}$ can be read out  from $D_2$. 

It can be verified directly that the Lagrangian \eqref{eq:discl3D} has the right symmetries. Namely, it is invariant under
\be\la{eq:symm}
\Phi^\prime = e^{i\sum_A c_A P_A(x)}\Phi\,,
\ee
and, consequently, has the right conservation laws and naturally couples to the multipole gauge theory \eqref{eq:HaahGauge} upon gauging \eqref{eq:symm}. 

The conservations laws are given by \eqref{eq:conservation} and imply that both $u-$ and $v-$ components of the dipole moment are conserved. Moreover, $Q_{uu}-Q_{vv}$ and $Q_{uv}$ components of the quadrupole tensor are also conserved. These conservation laws imply that charges, which correspond to disclinations, are mobile in $w$ direction, but immobile in $u$ and $v$ directions (quite similar to the $2$D case). While the dipoles, which correspond to dislocations with the Burgers vector along $w$ direction are planeons: due to quadrupole conservation laws, the dipoles can only move \emph{along} their dipole moments within the $u-v$ plane and they can move in $w$ direction. Note that in more familiar scalar charge theory the dipoles move \emph{perpendicular} to the dipole moment. 

Defect mobility analysis is in an agreement with what is known about smectic A phases \cite{kleman2007soft, kleman2008disclinations}. Namely, the dislocations are completely mobile and \emph{do not} satisfy the glide constraint. In fact, in smectics glide motion is \emph{harder} than the climb, while neither is prohibited by the conservations of ``layer number'' \cite{kleman2007soft}.  Dislocations are line defects, with the dislocation line pointing in the direction perpendicular to the dipole moment, \emph{within} the $u-v$ plane. The dislocation line appears to be invisible in the present dual theory, consequently we cannot distinguish between the edge and screw dislocations. In fact, the screw dislocations appear to not be available with the present formalism as they cannot be represented as a disclination dipole \cite{kamien2016topology, aharoni2017composite}. The motion of a dislocation along the defect line is not well-defined, this maps to dipole immobility in the direction perpendicular to the dipole moment. Disclinations can only move around $u-v$ plane by emitting or absorbing dislocations \cite{abukhdeir2008defect, kleman2008disclinations}. Finally, by comparing \eqref{eq:Gausslaw} and \eqref{eq:Efield} we can relate the defect density to the singularities in the derivatives of  the (singular part of) $\theta$. The defect density is $\rho \propto [D_1,D_2]\theta = [\p_w, \Delta_{u,v}]\theta$, which is a component of the disclination density \cite{Kleinert}. It's quite remarkable that after removing the information about dislocation line and ``modding out'' by dislocations loops the defects in smectic phases admit description in terms of (point) particles and dipoles. Of course, it is not possible, with the present formalism, to address the interesting questions of topology and geometry of dislocation links and knots.

%%%%%%%%%%%%%%%%%%%%
\paragraph{Conclusions.---} 
%%%%%%%%%%%%%%%%%%%%
We have described the map between the (continuum) $U(1)$ Haah code multipole gauge theory and an effective theory of a quantum $3$D smectic A phase. This map shows that multipole gauge theories can arise as dual descriptions of familiar phases. The seemingly obscure length scale dependence of the effective Lagrangian is very natural from the elastic point of view: smectic A phases consist of a set of two dimensional fluid layers stacked periodically in the third direction, with a fixed layer spacing that enters the Lagrangian explicitly. We have also exhibited a duality between a $2$D smectic phase and a multipole gauge theory. The multipole gauge theories are quite obscure objects, while spatially ordered quantum phases are significantly more familiar and are pretty well understood. We expect that systematic exploration of MGT dual to spatially ordered quantum mesophases will lead to the discovery of many new gauge theories and ultimately to new fracton phases. We also hope to apply these dualities to quantum Hall liquid crystal phases.

%%%%%%%%%%
\paragraph{Acknowledgments.---}
%%%%%%%%%%

It is a pleasure to thank R. Pelcovits and A. Souslov for patiently explaining to me the basic physics of smectic A phases. A.G. was supported by the Brown startup funds.

%%%%%%%%%%
\paragraph{Note added.---}
%%%%%%%%%%

Recently we became aware of a related work in progress on a duality between fractons and $2D$ quantum smectics \cite{2020Zhai}. 

%%%%%%%%%%
%%%%%%%%%%

\bibliography{Bibliography}

\newpage

\end{document}